\documentclass{ifacconf}

\usepackage{graphicx}      %
\usepackage{natbib}        %
\usepackage{underscore}
\usepackage{bm}
\usepackage{amssymb}
\usepackage[thinc]{esdiff}
\usepackage{tikz}
\usepackage[acronym]{glossaries}
\usepackage{pgfplotstable,pgfplots}
\usepackage{siunitx}
\usepackage{multirow}
\usepackage{multicol}
\usepackage{tabularx}
\pgfplotsset{compat=1.18}
\usetikzlibrary{positioning}
\usetikzlibrary {arrows.meta}
\usetikzlibrary{fit}
\usetikzlibrary{backgrounds}
\usepgfplotslibrary{units}
\usetikzlibrary{external}
\newcommand{\comment}[1]{}

\newcommand*{\pathVariableChar}{s}
\newcommand*{\pathChar}{\bm{r}}
\newcommand*{\horizontalPositionChar}{x}
\newcommand*{\verticalPositionChar}{y}

\newcommand*{\simulationTimestepDiscreteChar}{k}
\newcommand*{\orientationChar}{\theta}
\newcommand*{\curvatureChar}{\kappa}
\newcommand*{\frenetLateralOffsetChar}{d}
\newcommand*{\stateVectorChar}{\bm{x}}
\newcommand*{\inputVectorChar}{u}
\newcommand*{\disturbanceVectorChar}{z}
\newcommand*{\systemMatrixChar}{\bm{A}}
\newcommand*{\inputMatrixChar}{\bm{B}}
\newcommand*{\disturbanceMatrixChar}{\bm{D}}
\newcommand*{\sampleTimeChar}{T_s}
\newcommand*{\velocityChar}{v}
\newcommand*{\mpcStepChar}{\tau}
\newcommand*{\mpcHorizonChar}{N}
\newcommand*{\stateCostChar}{\bm{Q}}
\newcommand*{\controlCostChar}{R}
\newcommand*{\simulationHorizonChar}{M}
\newcommand*{\offsetChar}{\Delta}
\newcommand*{\arbitraryIndexCharA}{i}

\newcommand*{\frenetPointChar}{\bm{p}}

\newcommand*{\mpcCostWeightChar}{w}
\newcommand*{\costParameterVectorChar}{\bm{w}}
\newcommand*{\costParameterMatrixChar}{\bm{w}}
\newcommand*{\mpcCostWeightDecayChar}{\beta}

\newcommand*{\costFunctionChar}{J}
\newcommand*{\stateDimensionChar}{n}
\newcommand*{\noiseVectorSymbol}{\bm{\eta}}

\newcommand*{\referenceSubscript}{ref}
\newcommand*{\desiredSubscript}{des}
\newcommand*{\frenetPointFunctionSubscript}{pt}

\newcommand*{\simulationReferenceSuperscript}{sim}

\newcommand*{\R}{\mathbb{R}}
\newcommand*{\RPos}{\mathbb{R}_{\geq0}}
\newcommand*{\RstrictPos}{\mathbb{R}_{>0}}

\newcommand*{\diag}[1]{diag(#1)}
\DeclareMathOperator*{\argmin}{arg\,min}

\newcommand*{\arbitraryIndexA}{\arbitraryIndexCharA}

\newcommand*{\simulationTimestepDiscrete}{\simulationTimestepDiscreteChar}
\newcommand*{\sampleTime}{\sampleTimeChar}

\newcommand*{\stateDimension}{\stateDimensionChar}

\newcommand*{\vehicleHorizontalPosition}[1][\simulationTimestep]{\horizontalPositionChar(#1)}
\newcommand*{\vehicleVerticalPosition}[1][\simulationTimestep]{\verticalPositionChar(#1)}
\newcommand*{\vehicleVelocity}[1][\simulationTimestep]{\velocityChar(#1)}
\newcommand*{\vehicleVelocityDiscrete}[1][\simulationTimestepDiscrete]{\velocityChar_{#1}}
\newcommand*{\vehicleOrientation}[1][\simulationTimestep]{\orientationChar(#1)}
\newcommand*{\vehicleOrientationDiscrete}[1][\simulationTimestepDiscrete]{\orientationChar_{#1}}
\newcommand*{\vehicleCurvature}[1][\simulationTimestep]{\curvatureChar(#1)}
\newcommand*{\vehicleCurvatureDiscrete}[1][\simulationTimestepDiscrete]{\curvatureChar_{#1}}
\newcommand*{\vehicleCurvatureDot}[1][\simulationTimestep]{\dot{\curvatureChar}(#1)}
\newcommand*{\vehicleCurvatureDotDiscrete}[1][\simulationTimestepDiscrete]{\dot{\curvatureChar}_{#1}}
\newcommand*{\vehicleCurvatureDotDot}[1][\simulationTimestep]{\ddot{\curvatureChar}(#1)}
\newcommand*{\vehicleCurvatureDotDotDiscrete}[1][\simulationTimestepDiscrete]{\ddot{\curvatureChar}_{#1}}

\newcommand*{\referencePathVariableDiscreteSpecifiedReference}[2][\simulationTimestepDiscrete]{\pathVariableChar_{#1}^{(#2)}}
\newcommand*{\refPath}[1][\refPathVar]{\pathChar(#1)}

\newcommand*{\frenetControlPoint}[1][\simulationTimestep]{\pathVariableChar(#1)}
\newcommand*{\frenetLateralOffset}[1][\simulationTimestep]{\frenetLateralOffsetChar(#1)}
\newcommand*{\frenetLateralOffsetDiscrete}[1][\simulationTimestepDiscrete]{\frenetLateralOffsetChar_{#1}}
\newcommand*{\frenetReferencePathOrinetaion}[1][\simulationTimestep]{\orientationChar_\text{\referenceSubscript}(#1)}
\newcommand*{\frenetReferencePathOrinetaionDiscete}[1][\simulationTimestepDiscrete]{\orientationChar_{\text{\referenceSubscript},#1}}
\newcommand*{\frenetReferencePathOrinetaionDisceteSpecifiedReference}[2][\simulationTimestepDiscrete]{\orientationChar_{\text{\referenceSubscript},#1}^{(#2)}}
\newcommand*{\frenetReferencePathCurvature}[1][\simulationTimestep]{\curvatureChar_\text{\referenceSubscript}(#1)}

\newcommand*{\frenetReferencePathCurvatureDisceteSpecifiedReference}[2][\simulationTimestepDiscrete]{\curvatureChar_{\text{\referenceSubscript},#1}^{(#2)}}

\newcommand*{\frenetReferencePathCurvatureDotDiscreteSpecifiedReference}[2][\simulationTimestepDiscrete]{\dot{\curvatureChar}_{\text{\referenceSubscript},#1}^{(#2)}}

\newcommand*{\frenetStateVectorDiscrete}[1][\simulationTimestepDiscrete]{\stateVectorChar_{#1}}
\newcommand*{\frenetStateVectorDiscreteSpecifiedReference}[2][\simulationTimestepDiscrete]{\stateVectorChar_{#1}^{(#2)}}
\newcommand*{\frenetInputVectorDiscrete}[1][\simulationTimestepDiscrete]{\inputVectorChar_{#1}}
\newcommand*{\frenetDisturbanceVectorDiscrete}[1][\simulationTimestepDiscrete]{\disturbanceVectorChar_{#1}}
\newcommand*{\frenetSystemMatrixDiscrete}[1][\simulationTimestepDiscrete]{\systemMatrixChar_{#1}}
\newcommand*{\frenetInputMatrixDiscrete}[1][\simulationTimestepDiscrete]{\inputMatrixChar_{#1}}
\newcommand*{\frenetDisturbanceMatrixDiscrete}[1][\simulationTimestepDiscrete]{\disturbanceMatrixChar_{#1}}

\newcommand*{\mpcStep}{\mpcStepChar}
\newcommand*{\mpcHorizon}{\mpcHorizonChar}

\newcommand*{\mpcStateVector}[1][\simulationTimestepDiscrete+\mpcStep]{\hat{\stateVectorChar}_{#1}}
\NewDocumentCommand{\mpcDesiredStateVector}{ O{\simulationTimestepDiscrete+\mpcStep} O{\simulationTimestepDiscrete} }{\hat{\stateVectorChar}_{\text{\desiredSubscript},#1}^{(#2)}}
\newcommand*{\mpcInputVector}[1][\simulationTimestepDiscrete+\mpcStep]{\hat{\inputVectorChar}_{#1}}
\NewDocumentCommand{\mpcOptimizedInput}{ O{\simulationTimestepDiscrete+\mpcStep} O{\simulationTimestepDiscrete} }{\hat{\inputVectorChar}^{*(#2)}_{#1}}
\NewDocumentCommand{\mpcDisturbanceVector}{ O{\simulationTimestepDiscrete+\mpcStep} O{\simulationTimestepDiscrete} }{\hat{\disturbanceVectorChar}_{#1}^{(#2)}}
\NewDocumentCommand{\mpcSystemMatrix}{ O{\simulationTimestepDiscrete+\mpcStep} O{\simulationTimestepDiscrete} }{\hat{\systemMatrixChar}_{#1}^{(#2)}}
\NewDocumentCommand{\mpcInputMatrix}{ O{\simulationTimestepDiscrete+\mpcStep} O{\simulationTimestepDiscrete} }{\hat{\inputMatrixChar}_{#1}^{(#2)}}
\NewDocumentCommand{\mpcDisturbanceMatrix}{ O{\simulationTimestepDiscrete+\mpcStep} O{\simulationTimestepDiscrete} }{\hat{\disturbanceMatrixChar}_{#1}^{(#2)}}

\newcommand*{\mpcStateCost}[1][\mpcStep]{\hat{\stateCostChar}_{#1}}
\newcommand*{\mpcControlCost}[1][\mpcStep]{\hat{\controlCostChar}_{#1}}
\newcommand*{\mpcCostParameterVector}[1][\mpcStep]{\hat{\costParameterVectorChar}_{#1}}
\newcommand*{\mpcCostParameterMatrix}{\hat{\costParameterMatrixChar}}
\newcommand*{\mpcCostWeightLateralOffset}[1][\mpcStep]{\hat{\mpcCostWeightChar}_{\frenetLateralOffsetChar,#1}}
\newcommand*{\mpcCostWeightOrientation}[1][\mpcStep]{\hat{\mpcCostWeightChar}_{\orientationChar,#1}}
\newcommand*{\mpcCostWeightCurvature}[1][\mpcStep]{\hat{\mpcCostWeightChar}_{\curvatureChar0,#1}}
\newcommand*{\mpcCostWeightCurvatureDot}[1][\mpcStep]{\hat{\mpcCostWeightChar}_{\curvatureChar1,#1}}
\newcommand*{\mpcCostWeightCurvatureDotDot}[1][\mpcStep]{\hat{\mpcCostWeightChar}_{\curvatureChar2,#1}}
\newcommand*{\mpcCostWeightDecay}{\mpcCostWeightDecayChar}
\newcommand*{\mpcInputMin}{\inputVectorChar_{\text{min}}}
\newcommand*{\mpcInputMax}{\inputVectorChar_{\text{max}}}

\NewDocumentCommand{\mpcStateTrajectoryDiscrete}{ O{\simulationTimestepDiscrete} O{\simulationTimestepDiscrete+\mpcHorizon} }{\hat{\stateVectorChar}|_{#1}^{#2}}
\NewDocumentCommand{\mpcStateReferenceTrajectoryDiscrete}{ O{\simulationTimestepDiscrete} O{\simulationTimestepDiscrete+\mpcHorizon} O{\simulationTimestepDiscrete}}{\hat{\stateVectorChar}_{\text{\desiredSubscript}}^{(#3)}|_{#1}^{#2}}
\NewDocumentCommand{\mpcInputTrajectoryDiscrete}{ O{\simulationTimestepDiscrete} O{\simulationTimestepDiscrete+\mpcHorizon-1} }{\hat{\inputVectorChar}|_{#1}^{#2}}
\NewDocumentCommand{\mpcOptimizedInputTrajectoryDiscrete}{ O{\simulationTimestepDiscrete} O{\simulationTimestepDiscrete+\mpcHorizon-1} O{\simulationTimestepDiscrete}}{\hat{\inputVectorChar}^{*(#3)}|_{#1}^{#2}}
\NewDocumentCommand{\mpcReferenceNoise}{ O{\simulationTimestepDiscrete} O{\simulationTimestepDiscrete}}{\noiseVectorSymbol^{(#2)}_{#1}}

\newcommand*{\mpcCostFunctionPlain}{\hat{\costFunctionChar}_{\mpcCostParameterMatrix}}
\newcommand*{\mpcCostFunction}[1][\mpcStateTrajectoryDiscrete,\mpcStateReferenceTrajectoryDiscrete,\mpcInputTrajectoryDiscrete]{\hat{\costFunctionChar}_{\mpcCostParameterMatrix}(#1)}

\newcommand*{\simulationHorizon}{\simulationHorizonChar}
\newcommand*{\frenetOffsetStateVectorDiscrete}[3][\simulationTimestepDiscrete]{\offsetChar\stateVectorChar_{#1}^{(#2\rightarrow#3)}}
\newcommand*{\frenetOffsetLateralOffsetDiscrete}[3][\simulationTimestepDiscrete]{\offsetChar\frenetLateralOffsetChar_{#1}^{(#2\rightarrow#3)}}
\newcommand*{\frenetDesiredStateVectorDiscrete}[1][\simulationTimestepDiscrete]{\stateVectorChar_{\text{\desiredSubscript},#1}}
\newcommand*{\simulationStateCost}{\stateCostChar}
\newcommand*{\simulationControlCost}{\controlCostChar}
\newcommand*{\simulationCostParameterVector}{\costParameterVectorChar}
\newcommand*{\simulationCostWeightLateralOffset}{\mpcCostWeightChar_{\frenetLateralOffsetChar}}
\newcommand*{\simulationCostWeightOrientation}{\mpcCostWeightChar_{\orientationChar}}
\newcommand*{\simulationCostWeightCurvature}{\mpcCostWeightChar_{\curvatureChar0}}
\newcommand*{\simulationCostWeightCurvatureDot}{\mpcCostWeightChar_{\curvatureChar1}}
\newcommand*{\simulationCostWeightCurvatureDotDot}{\mpcCostWeightChar_{\curvatureChar2}}

\NewDocumentCommand{\simulationStateTrajectoryDiscrete}{ O{\simulationTimestepDiscrete} O{\simulationTimestepDiscrete+\simulationHorizon} }{\stateVectorChar|_{#1}^{#2}}
\NewDocumentCommand{\simulationStateReferenceTrajectoryDiscrete}{ O{\simulationTimestepDiscrete} O{\simulationTimestepDiscrete+\simulationHorizon} }{\stateVectorChar_\text{\desiredSubscript}|_{#1}^{#2}}
\NewDocumentCommand{\simulationInputTrajectoryDiscrete}{ O{\simulationTimestepDiscrete} O{\simulationTimestepDiscrete+\simulationHorizon-1} }{\inputVectorChar|_{#1}^{#2}}

\newcommand*{\simulationCostFunctionPlain}{\costFunctionChar_{\simulationCostParameterVector}}
\newcommand*{\simulationCostFunction}[1][{\simulationStateTrajectoryDiscrete},{\simulationStateReferenceTrajectoryDiscrete},{\simulationInputTrajectoryDiscrete}]{\costFunctionChar_{\simulationCostParameterVector}(#1)}

\newcommand*{\refPathSpecifiedReference}[2][\refPathVar]{\pathChar^{(#2)}(#1)}

\newcommand*{\frenetPoint}{\frenetPointChar}

\newcommand*{\frenetLateralOffsetFunctionSpecifiedReference}[2][\frenetPoint]{\frenetLateralOffsetChar_\text{\frenetPointFunctionSubscript}^{(#2)}(#1)}

\newacronym{adas}{ADAS}{Advanced Driver Assistance Systems}
\newacronym{tp}{TP}{Trajectory Planner}
\newacronym{qp}{QP}{Quadratic Programming}
\newacronym{mpc}{MPC}{Model Predictive Control}
\newacronym{lqr}{LQR}{Linear–Quadratic Regulator}
\newacronym{ioc}{IOC}{Inverse Optimal Control}
\newacronym{cfp}{CFP}{Cost Function Parameters}
\newacronym{dcfp}{DCFP}{Desired Cost Function Parameters}

\tikzset{
    >={Latex}
}
\usepackage{fancyhdr}
\newcommand{\mytitle}{\copyright 2025 the authors. This work has been accepted to IFAC for publication under a Creative Commons Licence CC-BY-NC-ND}
\begin{document}
\begin{frontmatter}

\title{Parameter Tuning Under Uncertain Road Perception in Driver Assistance Systems}%

\author[bmw]{Leon Greiser} 
\author[bmw]{Christian Rathgeber} 
\author[unibw]{Vladislav Nenchev}
\author[kit]{Sören Hohmann}

\address[bmw]{BMW Group, 85716 Unterschleissheim, Germany (e-mail: leon.greiser@bmw.de, christian.rathgeber@bmw.de).}
\address[unibw]{University of the Bundeswehr Munich, Werner-Heisenberg-Weg 39, 85579 Neubiberg, Germany (e-mail: vladislav.nenchev@unibw.de)}
\address[kit]{Institute of Control Systems, Karlsruhe Institute of Technology, 76131 Karlsruhe, Germany (e-mail: soeren.hohmann@kit.edu)}

\begin{abstract}                %
Advanced driver assistance systems have improved comfort, safety, and efficiency of modern vehicles. 
However, sensor limitations lead to noisy lane estimates that pose a significant challenge in developing performant control architectures. 
Lateral trajectory planning often employs an optimal control formulation to maintain lane position and minimize steering effort. 
The parameters are often tuned manually, which is a time-intensive procedure.
This paper presents an automatic parameter tuning method for lateral planning in lane-keeping scenarios based on recorded data, while taking into account noisy road estimates. 
By simulating the lateral vehicle behavior along a reference curve, our approach efficiently optimizes planner parameters for automated driving and demonstrates improved performance on previously unseen test data.
\end{abstract}


\end{frontmatter}
\thispagestyle{fancy}
\pagestyle{fancy}

\section{Introduction}\label{sec:introduction}

\Gls{adas} promise to improve driving comfort, safety, and energy efficiency~\citep{pradoOptimizingEnergyEfficientBraking2024}.
With the increasing number of \gls{adas} features, a unified underlying architecture is often used to reduce complexity and development costs.
This architecture typically includes an environment model, which estimates information relevant for the motion of the vehicle, such as the lane center.
Based on this, a \gls{tp} computes a motion trajectory for the vehicle.
The planning problem is typically formulated in \gls{mpc} fashion, where the deviation of the states from a reference, as well as the control input, is minimized over a defined time horizon.
It is then solved online in a receding horizon manner to adapt to changes in the environment.

Tuning the parameters of an \gls{mpc}-based \gls{tp} is often done manually by an expert to achieve the desired driving behavior.
The parameters, i.e. the cost function weights, depend not only on the specific vehicle but also on the scenario.
With \gls{adas} covering progressively more scenarios, this approach becomes too costly or even infeasible. 
In addition, \gls{mpc} has been reported to present challenges in tuning, mainly because one unified parameter set might not get the desired performance across different scenario clusters~\citep{zarroukiSafeReinforcementLearning2024}.

To save computational costs and simplify the design, the planning and following of the trajectory is often divided into a longitudinal and a lateral part~\citep{rajamaniVehicleDynamicsControl2012}.
This work focuses only on the lateral movement.
For lateral planning, the tracked reference is usually the lane center.
It is detected using a camera, sometimes in combination with map data.
As map data is not always available, accurate lane detection remains a challenging task.
Poor light and visibility conditions or degraded markings lead to inaccurate estimates of the lane center~\citep{dingLaneDetectionMethod2020}.
These inaccuracies, however, result in a suboptimal trajectory with respect to the true lane center.
To improve driving behavior, the noisy estimates have to be taken into account when tuning the parameters of the \gls{tp}.

This paper studies the tuning of the parameters of the existing \gls{tp}, while retaining its structure.
We propose a method to find a set of parameters for lane-keeping scenarios that, given an inaccurately estimated reference, leads as closely as possible to the desired driving behavior.
To this end, we make the following contributions.

\begin{enumerate}
    \item We derive a \gls{tp}-agnostic formulation of the parameter optimization problem.
    \item We optimize by re-simulating without noise distribution assumptions and utilize reference trajectories from recorded data instead.
    \item We demonstrate the generalization of the optimized parameters by showing an improved cost on test data.
\end{enumerate}

The remainder of the paper is organized as follows: In Section~\ref{sec:relatedWork} we discuss related work on tracking noisy references.
In Section~\ref{sec:preliminaries} we present the \gls{tp} formulation used, as well as the general problem statement.
In Section~\ref{sec:parameterTuning} we derive our parameter optimization approach.
The effectiveness of this approach is shown in Section~\ref{sec:experiments} using real-world data.
In Section~\ref{sec:conclusions} we present our conclusions.

\subsection{Related Work}\label{sec:relatedWork}

The availability of large datasets has driven the interest in data-driven approaches across many domains, including \gls{adas}.
A variety of \gls{tp} approaches have emerged that often rely on neural networks or other black-box architectures ~\citep{ganesanComprehensiveReviewDeep2024,redaPathPlanningAlgorithms2024}.
These approaches result in satisfactory performance in scenarios similar to those included in the training sets but lack explainability~\citep{tampuuSurveyEndtoEndDriving2022} and generalization~\citep{qureshiMotionPlanningNetworks2021} to unseen scenarios.

A common approach to deal with noise are robust control approaches such as tube-based~\citep{leeTubeBasedModelPredictive2024} or min-max~\citep{raimondoMinmaxModelPredictive2009} \gls{mpc}.
Although these approaches have strong stability guarantees under bounded noise, they can be conservative and lead to insufficient performance~\citep{liuDataDrivenDistributionallyRobust2023}. 
Unlike robust \gls{mpc}, we implicitly leverage the noise properties in the parameter optimization process through the training data.
If the probability distribution of the noise is known, stochastic \gls{mpc} can often be a better performing alternative~\citep{heirungStochasticModelPredictive2018}.
Probabilistic information on the road course can be integrated into the \gls{tp} using a target funnel to enhance steering behavior~\citep{bogenbergerTrajectoryPlanningAutomated2025}.
In this work, we assume no explicit knowledge of the noise distribution, a requirement for stochastic \gls{mpc}.

Another technique for dealing with reference signal noise is filtering with simple low-pass filters or more application-specific filters, such as a curvature corrected moving average~\citep{steineckerSimpleModelFreePath2023}.
Spline fitting and interpolation are especially common in lane detection systems~\citep{nuthongLaneDetectionUsing2010} as these approaches are well suited to remove high-frequency noise components.
In our experiments, we use reference data that has already been smoothed using spline fitting.

Due to computational restrictions and transparency, tuning is often preferred over advanced algorithms in practice~\citep{maciejowskiDiscussionMinmaxModel2009}.
Therefore, we focus on offline tuning of the cost parameters, leaving the controller structure itself unaltered.
\Citet{wuAutomaticParameterTuning2024} performed parameter optimization of the lateral \gls{tp} using different local optimization techniques and showed improved performance on the training data.
In contrast, we use a global optimization approach and show the generalization of the result by validation with test data.
Further, we provide a formulation as a single bilevel optimization problem, making the optimization applicable to different \glspl{tp}.

\section{Problem statement}\label{sec:preliminaries}

We introduce the state representation used for the \gls{tp}, followed by the \gls{tp} formulation.
Based on this, we present our problem statement.

\subsection{Vehicle Kinematics}

\begin{figure}
   \centering
   \includegraphics{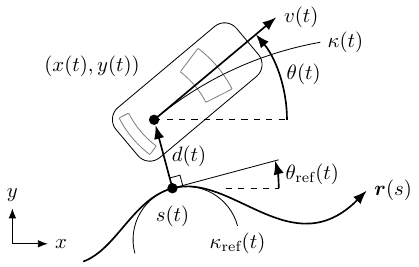}
   \caption{States with reference $\refPath$ \citep{gutjahrLateralVehicleTrajectory2016}.}
   \label{fig:frenet}
\end{figure}

The state of the vehicle is usually expressed in Cartesian coordinates as the position $(\vehicleHorizontalPosition, \vehicleVerticalPosition)$, the velocity $\vehicleVelocity$, the orientation $\vehicleOrientation$, and the driven curvature $\vehicleCurvature$.
Because embedded hardware in vehicles provides limited computing power, the kinematics are commonly linearized along a reference curve~\citep{gutjahrLateralVehicleTrajectory2016} as shown in Figure~\ref{fig:frenet}.
This allows the \gls{tp} to be divided into longitudinal and lateral parts.
Typically, the reference curve used is the lane center, expressed in Euclidean space as~$\refPath$.

The arc length of the reference curve to the point of the vehicle is~$\frenetControlPoint$.
The vector from that point to the vehicle is orthogonal to the reference curve. The length of this vector is equal to the absolute value of the lateral offset $\frenetLateralOffset$.
The orientation and curvature of the curve at that point are $\frenetReferencePathOrinetaion$ and $\frenetReferencePathCurvature$.
Using these coordinates, the kinematics of the lateral movements can be linearized and discretized~\citep{gutjahrLateralVehicleTrajectory2016} with sample time $\sampleTime$. 
The input is the second derivative of the curvature ${\frenetInputVectorDiscrete=\vehicleCurvatureDotDotDiscrete}$ to obtain a sufficiently smooth trajectory.
The orientation of the reference curve is treated as the disturbance ${\frenetDisturbanceVectorDiscrete=\frenetReferencePathOrinetaionDiscete}$ to the system. 
With the state vector $
    \frenetStateVectorDiscrete = 
[        \frenetLateralOffsetDiscrete, \vehicleOrientationDiscrete, \vehicleCurvatureDiscrete, \vehicleCurvatureDotDiscrete]^\top$
at time step $\simulationTimestepDiscrete$, the kinematics are

\begin{equation}\label{eq:stateSpace}
    \begin{aligned}
        \frenetStateVectorDiscrete[\simulationTimestepDiscrete+1]
        &=
        \underbrace{
            \left[\begin{array}{cccc}
                1 & \vehicleVelocityDiscrete\sampleTime & \frac{1}{2}\vehicleVelocityDiscrete^2\sampleTime^2 & \frac{1}{6}\vehicleVelocityDiscrete^2\sampleTime^3 \\
                0 & 1 & \vehicleVelocityDiscrete\sampleTime & \frac{1}{2}\vehicleVelocityDiscrete\sampleTime^2 \\
                0 & 0 & 1 & \sampleTime \\
                0 & 0 & 0 & 1 \\
            \end{array}\right]
        }_{\frenetSystemMatrixDiscrete}
        \frenetStateVectorDiscrete \\
        &+
        \underbrace{
            \left[\begin{array}{c}
                \frac{1}{24}\vehicleVelocityDiscrete^2\sampleTime^4 \\
                \frac{1}{6}\vehicleVelocityDiscrete\sampleTime^3 \\
                \frac{1}{2}\sampleTime^2 \\
                \sampleTime \\
            \end{array}\right]
        }_{\frenetInputMatrixDiscrete}
        \frenetInputVectorDiscrete
        +
        \underbrace{
            \left[\begin{array}{c}
                -\vehicleVelocityDiscrete\sampleTime \\
                0 \\
                0 \\
                0 \\
            \end{array}\right]
        }_{\frenetDisturbanceMatrixDiscrete}
        \frenetDisturbanceVectorDiscrete.
    \end{aligned}
\end{equation}

\subsection{Lateral Planning}

The goal of the \gls{tp} is to find a feasible and comfortable trajectory.
Therefore, the kinematics~\eqref{eq:stateSpace} are used to optimize the inputs over a limited time horizon in an \gls{mpc} fashion with the goal of minimizing the control input as well as the deviation of the states from a reference.
Boundary conditions for control input, safety, or comfort can be formulated as additional constraints.
The resulting optimization problem is a \gls{qp} problem that can be solved efficiently online.
In this work, we discard most of the boundary conditions for simplicity.
In this case, the trajectory planning problem at time step $\simulationTimestepDiscrete$ can be formulated using the cost function

\begin{equation}
    \mpcCostFunctionPlain = \textstyle\sum_{\mpcStep=0}^{\mpcHorizon} ||\mpcStateVector-\mpcDesiredStateVector||_{\mpcStateCost}^2 + \textstyle\sum_{\mpcStep=0}^{\mpcHorizon-1} ||\mpcInputVector||_{\mpcControlCost}^2
\end{equation}

as the optimization problem

\begin{equation}\label{eq:mpc}
    \begin{aligned}
        &\mpcOptimizedInputTrajectoryDiscrete = \\
        &\begin{aligned}[t]
            \argmin_{\substack{\mpcInputTrajectoryDiscrete}} \; & \mpcCostFunction \\
            \textrm{s.t.} \; & 
            \begin{aligned}[t]
                &\mpcStateVector[\simulationTimestepDiscrete+\mpcStep+1] = 
                \begin{aligned}[t]
                    &\mpcSystemMatrix \mpcStateVector + \mpcInputMatrix \mpcInputVector + \mpcDisturbanceMatrix \mpcDisturbanceVector
                \end{aligned} \\
                &\mpcStateVector[\simulationTimestepDiscrete] = \frenetStateVectorDiscreteSpecifiedReference{\simulationTimestepDiscrete}
            \end{aligned} \\
            & \mpcInputMin \leq \mpcInputVector \leq \mpcInputMax
        \end{aligned}
    \end{aligned}
\end{equation}

over the planning horizon of $\mpcHorizon$ time steps.
This formulation assumes that the longitudinal planning problem is already solved at time step $\simulationTimestepDiscrete$ and that the planned velocity, as well as the planned position along the reference curve, is known for each time step in the planning horizon.
The system matrix $\mpcSystemMatrix$, input matrix $\mpcInputMatrix$, and disturbance matrix $\mpcDisturbanceMatrix$ depend on the planned velocity.
The time step of planning is expressed in the superscript.
The expected disturbance $\mpcDisturbanceVector$ depends on the estimated reference curve and the planned positions along that curve.
Here, the superscript also expresses the time step of estimation of the reference curve.
The desired state $\mpcDesiredStateVector$ is derived from the reference curve as 

\begin{equation}\label{eq:referenceStateMpc}
    \mpcDesiredStateVector = \begin{bmatrix}
        0 & \frenetReferencePathOrinetaionDisceteSpecifiedReference[\simulationTimestepDiscrete+\mpcStep]{\simulationTimestepDiscrete} & \frenetReferencePathCurvatureDisceteSpecifiedReference[\simulationTimestepDiscrete+\mpcStep]{\simulationTimestepDiscrete} & \frenetReferencePathCurvatureDotDiscreteSpecifiedReference[\simulationTimestepDiscrete+\mpcStep]{\simulationTimestepDiscrete}
    \end{bmatrix}^\top.
\end{equation}

The state cost matrix ${\mpcStateCost \in \RPos^{4\times4}}$ and the control cost weight ${\mpcControlCost \in \RstrictPos}$ are defined as 

\begin{equation}\label{eq:mpcStateAndControlCostWeights}
    \begin{aligned}
        \mpcStateCost &= \diag{\begin{bmatrix}
            \mpcCostWeightLateralOffset & \mpcCostWeightOrientation & \mpcCostWeightCurvature & \mpcCostWeightCurvatureDot
        \end{bmatrix}^\top} \\
        \mpcControlCost &= \mpcCostWeightCurvatureDotDot.
    \end{aligned}
\end{equation}

For simplicity, we limit $\mpcStateCost$ to a diagonal matrix, although off-diagonal entries could be used.
Both can vary over the planning horizon and are parametrized by 

\begin{equation}\label{eq:mpcCostFunctionParameters}
    \begin{aligned}
        \mpcCostParameterVector &= \begin{bmatrix}
            \mpcCostWeightLateralOffset & \mpcCostWeightOrientation & \mpcCostWeightCurvature & \mpcCostWeightCurvatureDot & \mpcCostWeightCurvatureDotDot
        \end{bmatrix}^\top,\;\mpcCostParameterVector\in\RstrictPos^5\\
    \mpcCostParameterMatrix &= \begin{bmatrix}
        \mpcCostParameterVector[0]^\top & \mpcCostParameterVector[1]^\top & \dots & \mpcCostParameterVector[\mpcHorizon]^\top
    \end{bmatrix}^\top \text{ with } \mpcCostParameterVector = \mpcCostWeightDecay ^ \mpcStep \mpcCostParameterVector[0],
    \end{aligned}
\end{equation}

which we refer to as the \gls{cfp} in the following.
In this paper, we constrain the weights $\mpcCostParameterVector$ over the \gls{tp} horizon using the decay $\mpcCostWeightDecay \in [0,1]$ to reduce the number of parameters.
The state $\frenetStateVectorDiscreteSpecifiedReference{\arbitraryIndexA}$ at time step $\simulationTimestepDiscrete$ is relative to the reference curve estimated at time step $\arbitraryIndexA$.
Therefore, the initial planning state $\mpcStateVector[\simulationTimestepDiscrete] = \frenetStateVectorDiscreteSpecifiedReference{\simulationTimestepDiscrete}$ is the state at time step $\simulationTimestepDiscrete$ and relative to the reference curve estimated at time step $\simulationTimestepDiscrete$.

\subsection{Problem formulation}

We consider a discrete-time system 

\begin{equation}\label{eq:generalSystem}
    \frenetStateVectorDiscrete[\simulationTimestepDiscrete+1] = \frenetSystemMatrixDiscrete\frenetStateVectorDiscrete + \frenetInputMatrixDiscrete\frenetInputVectorDiscrete + \frenetDisturbanceMatrixDiscrete\frenetDisturbanceVectorDiscrete
\end{equation}

with the state vector $\frenetStateVectorDiscrete \in \R^\stateDimension$ and the input ${\frenetInputVectorDiscrete \in \R}$.
We aim to find a value for~$\mpcCostParameterMatrix$ of cost function~$\mpcCostFunctionPlain$ such that a defined cost $\simulationCostFunction$ over a horizon~$\simulationHorizon$ is minimal.
The state trajectory~$\simulationStateTrajectoryDiscrete$ is generated from the inputs~$\simulationInputTrajectoryDiscrete$ using~\eqref{eq:generalSystem}.
The inputs are generated by minimizing the cost function ${\mpcCostFunction}$ with~\eqref{eq:generalSystem} over ${\mpcHorizon < \simulationHorizon}$ in a receding horizon manner.
The noise~$\mpcReferenceNoise[\simulationTimestepDiscrete][\arbitraryIndexA]$, sampled at step~$\arbitraryIndexA$, of the desired state~${\mpcDesiredStateVector[\simulationTimestepDiscrete][\arbitraryIndexA]=\frenetDesiredStateVectorDiscrete+\mpcReferenceNoise[\simulationTimestepDiscrete][\arbitraryIndexA]}$ is of unknown distribution.
The goal is for~$\mpcCostParameterMatrix$ to compensate for noise that is state-specific and varies over the prediction horizon.

In the context of the lateral \gls{tp}, we aim to find a value for~$\mpcCostParameterMatrix$ of the \gls{tp} cost function~$\mpcCostFunctionPlain$ that minimizes the distance between the driven trajectory~$\simulationStateTrajectoryDiscrete$ and the true lane center~$\simulationStateReferenceTrajectoryDiscrete$ according to~$\simulationCostFunctionPlain$. 
To this end, we use recorded data of the inaccurately estimated lane center~$\mpcDesiredStateVector[\simulationTimestepDiscrete][\arbitraryIndexA]$ and the true lane center.

\section{Optimization-based Parameter Tuning}\label{sec:parameterTuning}

In this section, we derive a compact formulation for the discussed parameter optimization problem using a simulation of the kinematics with respect to a reference curve.
To this end, we present a method to solve the planning problem with a reference curve that is different from that used for the simulation.

Since we focus on the lateral motion, we also use~\eqref{eq:stateSpace} for simulation.
We assume that the tracking error of the planned trajectory is low in comparison to the inaccuracies of the estimated reference curve, as the vehicle dynamics are mostly compensated by a lower-level controller.
This way, we can focus on the \gls{tp} and close the loop using the linearized kinematics~\eqref{eq:stateSpace}.
This model depends on the velocity. 
Since only lateral motion is simulated, we assume that all longitudinal states are known \textit{a priori}.
For these, as well as for the reference curves, we used recorded data in our experiments in Section~\ref{sec:experiments}.

\subsection{Switching the Reference Curve of a State}

When simulating along a reference curve, states with different reference frames are used.
Besides the estimated reference curves for each time step $\simulationTimestepDiscrete$, i.e. the lane center, that are used to solve the \gls{mpc} problem, another reference curve is used for simulation.
This requires translating the simulation state into a new reference frame for the initial state of the \gls{mpc}.
The only part of the state vector $\frenetStateVectorDiscrete$ that depends on the reference curve is the lateral offset $\frenetLateralOffsetDiscrete$.
To transform the simulation state~$\frenetStateVectorDiscreteSpecifiedReference{\simulationReferenceSuperscript}$ to the initial state of the \gls{mpc}~$\frenetStateVectorDiscreteSpecifiedReference{\simulationTimestepDiscrete}$, only the lateral offset has to be transformed.
We define a state offset to switch the reference curve as 

\begin{equation}\label{eq:referenceSwitch}
    \begin{aligned}
        \frenetStateVectorDiscreteSpecifiedReference{\simulationTimestepDiscrete} &= \frenetStateVectorDiscreteSpecifiedReference{\simulationReferenceSuperscript} + \frenetOffsetStateVectorDiscrete{\simulationReferenceSuperscript}{\simulationTimestepDiscrete} \\
        \frenetOffsetStateVectorDiscrete{\simulationReferenceSuperscript}{\simulationTimestepDiscrete} &= 
        \begin{bmatrix}
            \frenetOffsetLateralOffsetDiscrete{\simulationReferenceSuperscript}{\simulationTimestepDiscrete} & 0 & 0 & 0
        \end{bmatrix}^\top.
    \end{aligned}
\end{equation}

Under the assumption that both curves are sufficiently smooth and parallel, we estimate the change in lateral offset using the small-angle approximation as

\begin{equation}
    \frenetOffsetLateralOffsetDiscrete{\simulationReferenceSuperscript}{\simulationTimestepDiscrete} \approx \frenetLateralOffsetFunctionSpecifiedReference[{\refPathSpecifiedReference[\referencePathVariableDiscreteSpecifiedReference{\simulationReferenceSuperscript}]{\simulationReferenceSuperscript}}]{\simulationTimestepDiscrete}.
\end{equation}

Here, $\referencePathVariableDiscreteSpecifiedReference{\simulationReferenceSuperscript}$ is the arc length to the vehicle along the simulation reference curve~$\refPathSpecifiedReference[\cdot]{\simulationReferenceSuperscript}$ at time step~$\simulationTimestepDiscrete$.
For a point~$\frenetPoint$, $\frenetLateralOffsetFunctionSpecifiedReference{\simulationTimestepDiscrete}$ is the lateral offset of that point from the reference curve estimated at time step~$\simulationTimestepDiscrete$.
Hence, $\frenetLateralOffsetFunctionSpecifiedReference[{\refPathSpecifiedReference[\referencePathVariableDiscreteSpecifiedReference{\simulationReferenceSuperscript}]{\simulationReferenceSuperscript}}]{\simulationTimestepDiscrete}$ is the lateral offset of the current point on the simulation reference curve relative to the reference curve estimated at time step~$\simulationTimestepDiscrete$.
With the reference curve used for simulation known \textit{a priori} and the longitudinal data, as well as the estimated reference curves obtained from real-world recordings, $\frenetOffsetLateralOffsetDiscrete{\simulationReferenceSuperscript}{\simulationTimestepDiscrete}$ can be calculated in advance of the optimization.

\subsection{Optimization Problem Formulation}

Equation~\eqref{eq:referenceSwitch} can be used to simulate the lateral vehicle movement along the true reference curve and solve the planning problem at each time step along the reference curve estimated at that time.
This allows us to formulate the parameter optimization problem as

\begin{equation}\label{eq:optimizationProblem}
    \begin{aligned}
        \min_{\mpcCostParameterVector[0],\mpcCostWeightDecay} \; & \simulationCostFunction \\
        \textrm{s.t.} \; & 
        \begin{aligned}[t]
            &\frenetStateVectorDiscrete[\simulationTimestepDiscrete+1] = \frenetSystemMatrixDiscrete \frenetStateVectorDiscrete + \frenetInputMatrixDiscrete \mpcOptimizedInput[\simulationTimestepDiscrete] + \frenetDisturbanceMatrixDiscrete \frenetDisturbanceVectorDiscrete \\
            &\eqref{eq:mpc}, \eqref{eq:referenceSwitch} \text{ with }
            \frenetStateVectorDiscreteSpecifiedReference{\simulationReferenceSuperscript} = \frenetStateVectorDiscrete
        \end{aligned}
    \end{aligned}
\end{equation}

using the simulation cost function

\begin{equation}
    \simulationCostFunctionPlain = \textstyle\sum_{\simulationTimestepDiscrete=0}^{\simulationHorizon} ||\frenetStateVectorDiscrete-\frenetDesiredStateVectorDiscrete||_{\simulationStateCost}^2 + \textstyle\sum_{\simulationTimestepDiscrete=0}^{\simulationHorizon-1} ||\mpcOptimizedInput[\simulationTimestepDiscrete]||_{\simulationControlCost}^2.
\end{equation}

The resulting problem is a bilevel optimization problem and is visualized in Figure~\ref{fig:title}.
The upper-level optimization task is to minimize the distance between the simulated states $\frenetStateVectorDiscrete$ and the desired states $\frenetDesiredStateVectorDiscrete$. 
The desired states are chosen analogously to~\eqref{eq:referenceStateMpc} from the simulation reference curve. 
Since this work focuses on the lane-keeping scenario, we choose the true lane center as the simulation reference curve.
For the lower level, there are multiple optimization tasks.
Equation~\eqref{eq:mpc} has to be solved for every time step $\simulationTimestepDiscrete$ in the simulation horizon $\simulationHorizon$.
The optimization variables are the \gls{mpc} \gls{cfp} ${\mpcCostParameterVector[0],\mpcCostWeightDecay}$ used in~\eqref{eq:mpcCostFunctionParameters}.

For the upper-level optimization problem, we use the diagonal state cost matrix ${\simulationStateCost \in \RPos^{4\times4}}$ and the control cost weight ${\simulationControlCost \in \RstrictPos}$

\begin{equation}
    \begin{aligned}
        \simulationStateCost &= \diag{\begin{bmatrix}
            \simulationCostWeightLateralOffset & \simulationCostWeightOrientation & \simulationCostWeightCurvature & \simulationCostWeightCurvatureDot
        \end{bmatrix}^\top} \\
        \simulationControlCost &= \simulationCostWeightCurvatureDotDot.
    \end{aligned}
\end{equation}

These define the desired behavior with regard to the true reference curve and can be chosen arbitrarily in reasonable bounds.
Given that they are defined for a finite receding horizon, we assume this horizon to be sufficiently large, such that the behavior approximately matches a trajectory optimized over the $\simulationHorizon$ steps, as done here.
$\simulationStateCost$ and $\simulationControlCost$ are parametrized by the vector

\begin{equation}
    \simulationCostParameterVector = \begin{bmatrix}
        \simulationCostWeightLateralOffset & \simulationCostWeightOrientation & \simulationCostWeightCurvature & \simulationCostWeightCurvatureDot & \simulationCostWeightCurvatureDotDot
    \end{bmatrix}^\top,\;\simulationCostParameterVector\in\RstrictPos^5,
\end{equation}

which will be referred to as the \gls{dcfp} in the following.

The lower-level tasks of~\eqref{eq:optimizationProblem} are convex \glspl{qp} and can be solved in polynomial time.
The upper-level optimization task is non-convex as it contains non-convex constraints.

\definecolor{colVar}{HTML}{33a02c}
\colorlet{colVarLight}{colVar!10}
\definecolor{colData}{HTML}{e31a1c}
\colorlet{colDataLight}{colData!10}
\definecolor{colParam}{HTML}{1f78b4}
\colorlet{colParamLight}{colParam!10}
\begin{figure}
   \centering
   \includegraphics{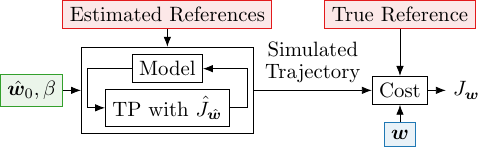}
   \caption{Block diagram of our bilevel optimization problem formulation. The parameters~{\color{colVar} (green)} of the \gls{tp} are tuned such that the simulated trajectory minimizes $\simulationCostFunctionPlain$. Reference trajectories~{\color{colData} (red)} and \gls{dcfp}~{\color{colParam} (blue)} are inputs to the optimization.}
   \label{fig:title}
\end{figure}

\section{Experimental Evaluation}\label{sec:experiments}

We demonstrate the effectiveness of tuning the \gls{cfp} in simulation, using the optimization problem formulation presented.
We optimize multiple sets of \gls{cfp} on training data, starting with arbitrarily chosen \gls{dcfp} $\simulationCostParameterVector$.
We compare the simulation cost using a \gls{tp} with the optimized \gls{cfp} and a \gls{tp} with the unaltered \gls{dcfp} on a test dataset.
In Section~\ref{sec:implementation} we first provide details on the implementation and data used.
In Section~\ref{sec:results} we present our results, which we discuss in Section~\ref{sec:discussion}.

\subsection{Implementation}\label{sec:implementation}

As described in the previous sections, we used data collected on a real-world vehicle.
For the reference curve of the \gls{tp}, we used the lane center of the online estimated road model, which includes a spline fitting.
In practice, it is difficult to measure the true lane center. 
In order to get a good estimate, the vehicle was driven in the center of the lane and the driven trajectory was extracted from the recorded odometry.
This data does not include the curvature derivative $\vehicleCurvatureDot$.
It was estimated using a Kalman filter followed by a Rauch-Tung-Striebel filter.
For the longitudinal data, e.g. the planned and simulated velocity, the recorded trajectory was used as well.
All data was interpolated and resampled to ${\sampleTime=\SI{0.1}{\second}}$.
A planning horizon of ${\mpcHorizon=30}$ steps was used.

The dataset we used has a total length of \SI{927}{\second}.
It was recorded on various types of roads with varying curvatures and conditions.
The velocity ranges from \SI{40}{\kilo\meter\per\hour} to \SI{100}{\kilo\meter\per\hour}.
For training we used \SI{795}{\second} and for testing \SI{132}{\second} of data.
The data consists of multiple continuous sections that vary in length from \SI{6}{\second} to \SI{60}{\second}.
Training and test data were split such that their state distributions match.

There are multiple methods to solve a bilevel optimization problem.
We used a nested evolutionary algorithm, which is a popular approach.
We solved the lower-level problems using the OSQP solver~\citep{stellatoOSQPOperatorSplitting2020}.
For the upper-level problem, we used differential evolution, a gradient-free heuristic~\citep{stornDifferentialEvolutionSimple1997}.

Because any set of cost function weights can be arbitrarily scaled without affecting the control law, we set $\mpcCostWeightCurvatureDotDot[0]=1$.
This leaves $\mpcCostWeightLateralOffset[0]$, $\mpcCostWeightOrientation[0]$, $\mpcCostWeightCurvature[0]$, $\mpcCostWeightCurvatureDot[0]$, and $\mpcCostWeightDecay$ as optimization variables.
We initialized $\mpcCostWeightDecay=1$ and the remaining optimization variables with the \gls{dcfp}.
We limited $\mpcCostParameterVector[0] \in [10^{-8}, 10^8]$ for numerical stability and $\mpcCostWeightDecay \in [0.5, 1]$.
To evaluate a set of \gls{cfp}, we calculated the cost for each continuous section of the dataset and added them together.

\subsection{Results}\label{sec:results}

Table~\ref{tb:desiredCostFunctionWeights} shows the ten sets of \gls{dcfp} \emph{A}-\emph{J} that we used.
The \gls{dcfp} were randomly generated.
We used various reasonably chosen \gls{cfp} sets to obtain an estimate of the variance of the error between the simulated and the desired states.
We used the inverse of these variances as a set of neutral \gls{cfp}.
We then multiplied these neutral \gls{cfp} with pseudo-random factors of $[0.25, 4]$.

Table~\ref{tb:optimizedCostFunctionWeights} shows the optimized \gls{cfp}.
To improve comparability, we scaled each set to a magnitude similar to \gls{dcfp}.
It can be seen that for $\mpcCostWeightLateralOffset[0]$, $\mpcCostWeightCurvatureDot[0]$, and $\mpcCostWeightCurvatureDotDot[0]$, the weights have similar orders of magnitude.
In contrast, outliers with significantly lower values can be observed for $\mpcCostWeightOrientation[0]$ and $\mpcCostWeightCurvature[0]$.
Most of $\mpcCostWeightCurvatureDot[0]$ are higher than $\simulationCostWeightCurvatureDot$. 
This is because the derivative of the curvature has significantly more noise when estimated using a Kalman filter as opposed to the estimate from the fitted spline. 
For $\mpcCostWeightDecay$, all values are higher than $0.97$.
With this decay, the weights are reduced by \SI{60}{\percent} at the end of the \gls{tp} horizon.
Lower values would significantly shorten the effective length of the horizon.

Table~\ref{tb:testLoss} shows the simulation cost on the test dataset using the \gls{dcfp} and the optimized \gls{cfp} for the \gls{tp}.
The right column further shows the relative change with a negative value showing an improvement.
Except for \emph{E}, \emph{F}, and \emph{G}, all optimized \gls{cfp} show an improvement, with \emph{C} showing the most significant.
The average relative change across all ten sets is \SI{-4.13}{\percent}.

Figure~\ref{fig:simulationResults} shows the simulated trajectories for a part of the test data.
A \gls{tp} using the optimized \gls{cfp} and the \gls{dcfp} of set \emph{C} are compared.
The lateral offset~$\frenetLateralOffset$ using the optimized \gls{cfp} is closer to the desired value most of the time.
For the orientation~$\vehicleOrientation$, no large deviations can be observed.
In the case of the curvature~$\vehicleCurvature$ and curvature deviation~$\vehicleCurvatureDot$, the trajectory based on the optimized \gls{cfp} is smoother.
Further, in both cases, the desired states are being followed more accurately.
Lastly, in case of the control input~$\vehicleCurvatureDotDot$, a smaller amplitude can be observed, leading to the overall smoother trajectory.

The average absolute deviation of the states from their desired values over the test dataset are {\num{0.119}}, {\num{0.00586}}, {\num{0.591e-3}}, and {\num{1.33e-3}}.
The maximum absolute deviations are {\num{0.646}}, {\num{0.0472}}, {\num{6.59e-3}}, and {\num{11.9e-3}}.
The input values are constrained by the \gls{mpc}.
Although we do not perform a formal stability analysis, these values indicate that with the optimized \gls{mpc} parameters, the states remain close to their reference over the test dataset.

\begin{table}[hb]
    \begin{center}
        \caption{Desired Cost Function Parameters}
        \label{tb:desiredCostFunctionWeights}
        \begin{tabular}{c|ccccc}
            \multirow{2}{*}{\bf{Set}} & $\bm{\simulationCostWeightLateralOffset}$ & $\bm{\simulationCostWeightOrientation}$ & $\bm{\simulationCostWeightCurvature}$ & $\bm{\simulationCostWeightCurvatureDot}$ & $\bm{\simulationCostWeightCurvatureDotDot}$ \\
            & $\times 10^1$ & $\times 10^4$ & $\times 10^6$ & $\times 10^5$ & $\times 10^4$ \\\hline
            A & \num{1.49} & \num{3.38} & \num{1.61} & \num{2.34} & \num{0.846} \\
            B & \num{0.811} & \num{0.284} & \num{2.33} & \num{2.36} & \num{3.91} \\
            C & \num{0.557} & \num{3.56} & \num{2.13} & \num{0.803} & \num{0.908} \\
            D & \num{0.875} & \num{0.563} & \num{0.906} & \num{1.48} & \num{1.23} \\
            E & \num{2.87} & \num{0.356} & \num{0.475} & \num{1.23} & \num{1.94} \\
            F & \num{2.84} & \num{0.388} & \num{0.253} & \num{6.19} & \num{7.98} \\
            G & \num{0.738} & \num{0.956} & \num{0.233} & \num{5.55} & \num{1.12} \\
            H & \num{7.74} & \num{2.08} & \num{2.86} & \num{5.33} & \num{2.88} \\
            I & \num{1.55} & \num{0.514} & \num{2.10} & \num{1.20} & \num{1.20} \\
            J & \num{2.37} & \num{0.358} & \num{1.95} & \num{0.548} & \num{8.46} \\
        \end{tabular}
    \end{center}
\end{table}

\begin{table*}[ht]
	\begin{tabularx}{\linewidth}{@{}>{\hsize=1.1\hsize}X@{}@{}>{\hsize=0.9\hsize}X@{}}
		\begin{minipage}{\hsize}
			\begin{center}
                \caption{Optimized Cost Function Weights}
                \label{tb:optimizedCostFunctionWeights}
                \begin{tabular}{c|cccccc}
                    \multirow{2}{*}{\bf{Set}} & $\bm{\mpcCostWeightLateralOffset[0]}$ & $\bm{\mpcCostWeightOrientation[0]}$ & $\bm{\mpcCostWeightCurvature[0]}$ & $\bm{\mpcCostWeightCurvatureDot[0]}$ & $\bm{\mpcCostWeightCurvatureDotDot[0]}$ & \multirow{2}{*}{$\bm{\mpcCostWeightDecay}$} \\
                    & $\times 10^1$ & $\times 10^4$ & $\times 10^6$ & $\times 10^5$ & $\times 10^4$ & \\\hline
                    A & \num{1.96} & \num{0.442} & \num{2.58e-06} & \num{6.33} & \num{0.143} & \num{1.00} \\
                    B & \num{0.821} & \num{0.150} & \num{0.278} & \num{6.16} & \num{1.47} & \num{1.00} \\
                    C & \num{1.29} & \num{0.813} & \num{7.61e-06} & \num{6.28} & \num{0.0637} & \num{0.992} \\
                    D & \num{1.78} & \num{0.196} & \num{0.202} & \num{6.02} & \num{0.742} & \num{0.997} \\
                    E & \num{3.53} & \num{2.87e-05} & \num{0.126} & \num{4.56} & \num{1.34} & \num{0.976} \\
                    F & \num{1.96} & \num{2.91e-04} & \num{0.228} & \num{4.28} & \num{3.07} & \num{0.974} \\
                    G & \num{1.34} & \num{0.129} & \num{0.248} & \num{6.76} & \num{0.327} & \num{0.998} \\
                    H & \num{2.92} & \num{0.156} & \num{1.84e-10} & \num{5.48} & \num{0.841} & \num{0.997} \\
                    I & \num{1.89} & \num{1.58e-07} & \num{0.434} & \num{6.22} & \num{0.230} & \num{1.00} \\
                    J & \num{1.58} & \num{0.0152} & \num{0.196} & \num{3.74} & \num{4.18} & \num{1.00} \\
                \end{tabular}
            \end{center}
		\end{minipage} &
        \begin{minipage}{\hsize}
			\begin{center}
                \caption{Cost~$\simulationCostFunctionPlain$ on Test Dataset}
                \label{tb:testLoss}
                \begin{tabular}{c|ccc}
                    \multirow{2}{*}{\bf{Set}} & \bf{\gls{tp} with} & \bf{\gls{tp} with} & \bf{Relative} \\
                     & \bf{\gls{dcfp}} & \bf{optimized \gls{cfp}} & \bf{Change} \\\hline
                    A & \num{7135.3} & \num{6480.6} & \SI{-9.18}{\percent} \\
                    B & \num{5621.0} & \num{5386.5} & \SI{-4.17}{\percent} \\
                    C & \num{6923.8} & \num{5863.5} & \SI{-15.32}{\percent} \\
                    D & \num{3047.7} & \num{2968.7} & \SI{-2.59}{\percent} \\
                    E & \num{3040.3} & \num{3066.5} & +\SI{0.86}{\percent} \\
                    F & \num{6810.1} & \num{6870.7} & +\SI{0.89}{\percent} \\
                    G & \num{4622.1} & \num{4688.5} & +\SI{1.44}{\percent} \\
                    H & \num{11622} & \num{11589} & \SI{-0.28}{\percent} \\
                    I & \num{4530.8} & \num{4266.6} & \SI{-5.83}{\percent} \\
                    J & \num{6249.8} & \num{5806.9} & \SI{-7.09}{\percent} \\
                \end{tabular}
            \end{center}
		\end{minipage}
	\end{tabularx}
\end{table*}

\definecolor{demonstrationColor}{HTML}{1f78b4}  %
\definecolor{optimizedColor}{HTML}{33a02c}  %
\definecolor{unoptimizedColor}{HTML}{e31a1c}  %
\begin{figure}
    \centering%
    \includegraphics{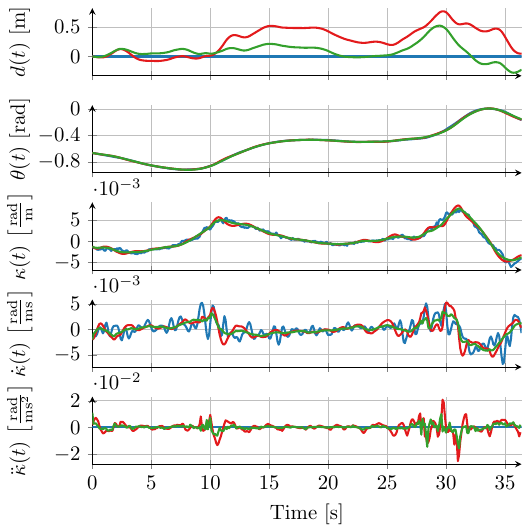}
        \caption{Simulation on part of the test data using the \gls{tp} with optimized \gls{cfp} {\color{optimizedColor}(green)} \emph{C} and with \gls{dcfp} {\color{unoptimizedColor}(red)}. The desired trajectory {\color{demonstrationColor}(blue)}, i.e. lane center, is included for reference.}
        \label{fig:simulationResults}
\end{figure}

\subsection{Discussion}\label{sec:discussion}

The results in Section~\ref{sec:results} show that for most of the chosen \gls{dcfp}, the optimized \gls{cfp} show improved performance over the \gls{dcfp} when used in a \gls{tp} with an inaccurate reference.
While a cost reduction can be expected on the training dataset, we also showed a cost reduction on the test dataset.
This indicates a generally better performance of the optimized \gls{cfp} in scenarios covered by the test data.
Further, we have shown that the optimization of the \gls{cfp} works for a variety of desired cost functions if they are chosen in reasonable bounds.
However, for a deeper understanding of the relationship between the desired cost function and performance gain, more \gls{dcfp} sets would have to be evaluated.
If the assumptions posed in this work hold, these improvements can be expected to transfer to the real-world.
We matched ${\sampleTime}$ and ${\mpcHorizon}$ to the real system in this work, although variations may affect performance.
While we provide a value for the average cost improvement, its impact on perceived driving dynamics is unclear.
With an average optimization runtime of \SI{2.5}{\hour} this approach is faster than manual tuning, although it is not yet scalable to large amounts of data.

\section{Conclusions}\label{sec:conclusions}

In this work, we studied the optimization of the parameters of the cost function of an optimal controller to compensate for inaccuracies in its reference trajectory.
Specifically, we investigated this problem for the lateral trajectory planning of a vehicle in a lane-keeping scenario, with the lane center as the reference.
We derived a compact bilevel optimization problem formulation for the lateral movement of the vehicle to reduce the value of the desired cost function with respect to the true reference.
Using collected real-world data, we performed multiple optimizations with different desired cost functions.
With a reduced cost on the test dataset, we showed a general improvement toward the desired driving behavior in simulation.
The main advantage of our approach is its practical relevance and simplicity.
Instead of adding complexity by modifying the controller itself, it leverages offline optimization on collected data.

\bibliography{references}

\begin{thebibliography}{20}
\providecommand{\natexlab}[1]{#1}
\providecommand{\url}[1]{\texttt{#1}}
\providecommand{\urlprefix}{URL }
\expandafter\ifx\csname urlstyle\endcsname\relax
  \providecommand{\doi}[1]{doi:\discretionary{}{}{}#1}\else
  \providecommand{\doi}{doi:\discretionary{}{}{}\begingroup \urlstyle{rm}\Url}\fi

\bibitem[{Bogenberger et~al.(2025)Bogenberger, B{\"u}rger, and Nenchev}]{bogenbergerTrajectoryPlanningAutomated2025}
Bogenberger, B., B{\"u}rger, J., and Nenchev, V. (2025).
\newblock Trajectory {{Planning}} for {{Automated Driving}} using {{Target Funnels}}.
\newblock In \emph{2025 {{European Control Conference}} ({{ECC}})}. IEEE, Thessaloniki, Greece.

\bibitem[{Ding et~al.(2020)Ding, Zhang, Xiao, Shu, and Lu}]{dingLaneDetectionMethod2020}
Ding, L., Zhang, H., Xiao, J., Shu, C., and Lu, S. (2020).
\newblock A {{Lane Detection Method Based}} on {{Semantic Segmentation}}.
\newblock \emph{Computer Modeling in Engineering \& Sciences}, 122(3), 1039--1053.

\bibitem[{Ganesan et~al.(2024)Ganesan, Kandhasamy, Chokkalingam, and {Mihet-Popa}}]{ganesanComprehensiveReviewDeep2024}
Ganesan, M., Kandhasamy, S., Chokkalingam, B., and {Mihet-Popa}, L. (2024).
\newblock A {{Comprehensive Review}} on {{Deep Learning-Based Motion Planning}} and {{End-to-End Learning}} for {{Self-Driving Vehicle}}.
\newblock \emph{IEEE Access}, 12, 66031--66067.

\bibitem[{Gutjahr et~al.(2016)Gutjahr, Groll, and Werling}]{gutjahrLateralVehicleTrajectory2016}
Gutjahr, B., Groll, L., and Werling, M. (2016).
\newblock Lateral {{Vehicle Trajectory Optimization Using Constrained Linear Time-Varying MPC}}.
\newblock \emph{IEEE Transactions on Intelligent Transportation Systems}, 18(6), 1586--1595.

\bibitem[{Heirung et~al.(2018)Heirung, Paulson, O'Leary, and Mesbah}]{heirungStochasticModelPredictive2018}
Heirung, T.A.N., Paulson, J.A., O'Leary, J., and Mesbah, A. (2018).
\newblock Stochastic model predictive control --- how does it work?
\newblock \emph{Computers \& Chemical Engineering}, 114, 158--170.

\bibitem[{Lee and Jeong(2024)}]{leeTubeBasedModelPredictive2024}
Lee, T. and Jeong, Y. (2024).
\newblock A {{Tube-Based Model Predictive Control}} for {{Path Tracking}} of {{Autonomous Articulated Vehicle}}.
\newblock \emph{Actuators}, 13(5), 164.

\bibitem[{Liu et~al.(2023)Liu, Shi, and Tokekar}]{liuDataDrivenDistributionallyRobust2023}
Liu, R., Shi, G., and Tokekar, P. (2023).
\newblock Data-{{Driven Distributionally Robust Optimal Control}} with {{State-Dependent Noise}}.
\newblock In \emph{2023 {{IEEE}}/{{RSJ International Conference}} on {{Intelligent Robots}} and {{Systems}} ({{IROS}})}, 9986--9991. IEEE, Detroit, MI, USA.

\bibitem[{Maciejowski(2009)}]{maciejowskiDiscussionMinmaxModel2009}
Maciejowski, J. (2009).
\newblock Discussion on: ``{{Min-max Model Predictive Control}} of {{Nonlinear Systems}}: {{A Unifying Overview}} on {{Stability}}''.
\newblock \emph{European Journal of Control}, 15(1), 22--25.

\bibitem[{Nuthong and Charoenpong(2010)}]{nuthongLaneDetectionUsing2010}
Nuthong, C. and Charoenpong, T. (2010).
\newblock Lane detection using smoothing spline.
\newblock In \emph{2010 3rd {{International Congress}} on {{Image}} and {{Signal Processing}}}, 989--993. IEEE, Yantai, China.

\bibitem[{Prado et~al.(2024)Prado, Nenchev, and Rathgeber}]{pradoOptimizingEnergyEfficientBraking2024}
Prado, A.A., Nenchev, V., and Rathgeber, C. (2024).
\newblock Optimizing {{Energy-Efficient Braking Trajectories}} with {{Anticipatory Road Data}} for {{Automated Vehicles}}.
\newblock In \emph{2024 {{European Control Conference}} ({{ECC}})}, 3280--3285. IEEE, Stockholm, Sweden.

\bibitem[{Qureshi et~al.(2021)Qureshi, Miao, Simeonov, and Yip}]{qureshiMotionPlanningNetworks2021}
Qureshi, A.H., Miao, Y., Simeonov, A., and Yip, M.C. (2021).
\newblock Motion {{Planning Networks}}: {{Bridging}} the {{Gap Between Learning-Based}} and {{Classical Motion Planners}}.
\newblock \emph{IEEE Transactions on Robotics}, 37(1), 48--66.

\bibitem[{Raimondo et~al.(2009)Raimondo, Limon, Lazar, Magni, and Ndez~Camacho}]{raimondoMinmaxModelPredictive2009}
Raimondo, D.M., Limon, D., Lazar, M., Magni, L., and Ndez~Camacho, E.F. (2009).
\newblock Min-max {{Model Predictive Control}} of {{Nonlinear Systems}}: {{A Unifying Overview}} on {{Stability}}.
\newblock \emph{European Journal of Control}, 15(1), 5--21.

\bibitem[{Rajamani(2012)}]{rajamaniVehicleDynamicsControl2012}
Rajamani, R. (2012).
\newblock \emph{Vehicle {{Dynamics}} and {{Control}}}.
\newblock Mechanical {{Engineering Series}}. Springer US, Boston, MA.

\bibitem[{Reda et~al.(2024)Reda, Onsy, Haikal, and Ghanbari}]{redaPathPlanningAlgorithms2024}
Reda, M., Onsy, A., Haikal, A.Y., and Ghanbari, A. (2024).
\newblock Path planning algorithms in the autonomous driving system: {{A}} comprehensive review.
\newblock \emph{Robotics and Autonomous Systems}, 174, 104630.

\bibitem[{Steinecker and Wuensche(2023)}]{steineckerSimpleModelFreePath2023}
Steinecker, T. and Wuensche, H.J. (2023).
\newblock A {{Simple}} and {{Model-Free Path Filtering Algorithm}} for {{Smoothing}} and {{Accuracy}}.
\newblock In \emph{2023 {{IEEE Intelligent Vehicles Symposium}} ({{IV}})}, 1--7. IEEE, Anchorage, AK, USA.

\bibitem[{Stellato et~al.(2020)Stellato, Banjac, Goulart, Bemporad, and Boyd}]{stellatoOSQPOperatorSplitting2020}
Stellato, B., Banjac, G., Goulart, P., Bemporad, A., and Boyd, S. (2020).
\newblock {{OSQP}}: An operator splitting solver for quadratic programs.
\newblock \emph{Mathematical Programming Computation}, 12(4), 637--672.

\bibitem[{Storn and Price(1997)}]{stornDifferentialEvolutionSimple1997}
Storn, R. and Price, K. (1997).
\newblock Differential {{Evolution}} -- {{A Simple}} and {{Efficient Heuristic}} for global {{Optimization}} over {{Continuous Spaces}}.
\newblock \emph{Journal of Global Optimization}, 11(4), 341--359.

\bibitem[{Tampuu et~al.(2022)Tampuu, Matiisen, Semikin, Fishman, and Muhammad}]{tampuuSurveyEndtoEndDriving2022}
Tampuu, A., Matiisen, T., Semikin, M., Fishman, D., and Muhammad, N. (2022).
\newblock A {{Survey}} of {{End-to-End Driving}}: {{Architectures}} and {{Training Methods}}.
\newblock \emph{IEEE Transactions on Neural Networks and Learning Systems}, 33(4), 1364--1384.

\bibitem[{Wu et~al.(2024)Wu, Nenchev, and Rathgeber}]{wuAutomaticParameterTuning2024}
Wu, H.J., Nenchev, V., and Rathgeber, C. (2024).
\newblock Automatic {{Parameter Tuning}} of {{Self-Driving Vehicles}}.
\newblock In \emph{2024 {{IEEE Conference}} on {{Control Technology}} and {{Applications}} ({{CCTA}})}, 555--560. IEEE, Newcastle upon Tyne, United Kingdom.

\bibitem[{Zarrouki et~al.(2024)Zarrouki, Spanakakis, and Betz}]{zarroukiSafeReinforcementLearning2024}
Zarrouki, B., Spanakakis, M., and Betz, J. (2024).
\newblock A {{Safe Reinforcement Learning}} driven {{Weights-varying Model Predictive Control}} for {{Autonomous Vehicle Motion Control}}.
\newblock In \emph{2024 {{IEEE Intelligent Vehicles Symposium}} ({{IV}})}, 1401--1408. IEEE, Jeju Island, Korea, Republic of.

\end{thebibliography}

\end{document}